\documentclass[twocolumn,5p]{elsarticle}
\usepackage{amssymb}
\usepackage{lipsum,xcolor}
\usepackage[numbers]{natbib}
\usepackage{parskip}

\begin{document}
\begin{frontmatter}

\title{Exploring Superconductivity in \textcolor{black}{Ba$_{3}$Ir$_{4}$Ge$_{16}$}: Experimental and Theoretical Insights}

\author[1]{A. Bhattacharyya}
\ead{amitava.bhattacharyya@rkmvu.ac.in}
\address[1]{Ramakrishna Mission Vivekananda Educational and Research Institute, Howrah 711202, West Bengal, India}

\author[2,3]{D. T. Adroja}
\address[2]{ISIS Facility, Rutherford Appleton Laboratory, Chilton, Didcot, Oxon OX11 0QX, United Kingdom}
\address[3]{Highly Correlated Matter Research Group, Physics Department, University of Johannesburg, Auckland Park 2006, South Africa}

\author[1]{A. K. Jana}

\author[1,4]{K. Panda}
\address[4]{Department of Physics, Ariel University, Ariel 40700, Israel}

\author[5]{P.P. Ferreira}
\ead{pedroferreira@usp.br}
\address[5]{Computational Materials Science Group (ComputEEL/MatSci), Universidade de São Paulo, Escola de Engenharia de Lorena, DEMAR, Lorena, Brazil}

\author[6]{Y. Zhao}
\address[6]{School of Physical Science and Technology, ShanghaiTech University, Shanghai 201210, China}

\author[7]{T. Ying}
\address[7]{Beijing National Laboratory for Condensed Matter Physics, Institute of Physics, Chinese Academy of Sciences, Beijing 100190, China}

\author[8]{H. Hosono}
\address[8]{Materials Research Center for Element Strategy, Tokyo Institute of Technology, 4259 Nagatsuta, Midori-ku, Yokohama 226-8503, Japan}

\author[9]{T. T. Dorini}
\address[9]{Université de Lorraine, CNRS, IJL, Nancy, France}

\author[9]{L. T. F. Eleno}

\author[2]{P. K. Biswas\fnref{fn1}}

\author[2]{G. Stenning}

\author[2]{R. Tripathi}

\author[10]{Y. Qi }
\ead{qiyp@shanghaitech.edu.cn}
\address[10]{ShanghaiTech Laboratory for Topological Physics, ShanghaiTech University, Shanghai 201210, China}

\fntext[fn1]{Deceased}

\begin{abstract}

We explore both experimental and theoretical aspects of the superconducting properties in the distinctive layered caged compound, Ba$_{3}$Ir$_{4}$Ge$_{16}$. Our approach integrates muon spin rotation and relaxation ($\mu$SR) measurements with magnetization and heat capacity experiments, accompanied by first-principle calculations. The compound's bulk superconductivity is unequivocally established through DC magnetization measurements, revealing a critical temperature ($T_\mathrm{C}$) of 5.7 K. A noteworthy characteristic observed in the low-temperature superfluid density is its saturating behavior, aligning with the features typical of conventional Bardeen-Cooper-Schrieffer (BCS) superconductors. The assessment of moderate electron-phonon coupling superconductivity is conducted through transverse field $\mu$SR measurements, yielding a superconducting gap to $T_\mathrm{C}$ ratio ($2\Delta(0)/k_\mathrm{B}T_\mathrm{C}$) of 4.04, a value corroborated by heat capacity measurements. Crucially, zero field $\mu$SR measurements dismiss the possibility of any spontaneous magnetic field emergence below $T_\mathrm{C}$, highlighting the preservation of time-reversal symmetry. Our experimental results are reinforced by first-principles density functional calculations, underscoring the intricate interplay between crystal structure and superconducting order parameter symmetry in polyhedral caged compounds. This comprehensive investigation enhances our understanding of the nuanced relationship between crystal structure and superconductivity in such unique compounds.
\end{abstract}




\end{frontmatter}

\section{Introduction}
\label{introduction}

We explore the intricate relationship between the structural characteristics of materials and their superconducting properties, focusing on a distinctive class known for its low-dimensional structural units. Materials exhibiting 1-D chains, 2-D planes, ladders, and layered structures often showcase enhanced charge, spin, or orbital fluctuations, leading to unconventional superconductivity, as exemplified in well-known instances like iron pnictides and copper oxides~\cite{lee2006,stewart2011}. A particularly promising avenue of exploration lies in polyhedral caged structures. These materials, characterized by their structural flexibility and a diverse range of fascinating physical properties, have garnered substantial interest within the scientific community~\cite{pei2022pressure}.

Categorized by the position of the guest atom within their structure, caged superconductors are divided into two main groups. The first category involves alkali metals positioned outside football-shaped fullerenes (C$_{60}$), as observed in fulleride-based doped caged compounds~\cite{kelty1991,ellis2007caged,rosseinsky1991,hebard1991}. In the second category, guest atoms are strategically placed at the center of the cages, defining filled caged type compounds. Examples of these include $\beta$-pyrochlore oxides AOs$_{2}$O$_{6}$ (A = alkaline metals)~\cite{yonezawa2004,hiroi2011,hiroi2004,saniz2004,hiroi2005,muramatsu2004,yamaura2006,saniz2005}, clathrates B$_{8}$(Ge,Si)$_{46}$ (B = alkaline earth metals)~\cite{yamanaka2000,bouquet2001,aydemir2010}, and filled skutterudites AT$_{4}$X$_{12}$ (A = alkali metals, T = transition metals, X = pnictogens)~\cite{yamaura2006}. Recent attention has also been directed towards high-$T_\mathrm{C}$ superhydride superconductors, such as YH$_{9}$ and LaH$_{10}$~\cite{heil2019, drozdov2019,somayazulu2019,hong2020,snider2021}.

A key aspect in the emergence of superconductivity in these compounds lies in the interaction between host-guest and interframework elements. Local low-frequency anharmonic lattice vibrations, commonly referred to as rattling phonons, play a crucial role by forming pairs with the conduction electrons of the cage. Particularly noteworthy is the scenario where the cage dimensions exceed those of the guest atom placed at its center. In such instances, the electron-phonon interaction is significantly enhanced, thereby favoring the onset of spontaneous gauge-symmetry breaking instabilities. This study aims to shed light on the intricate interplay between crystal structure and superconductivity within these unique materials.

Within this context, Ba$_{3}$Rh$_{4}$Ge$_{16}$, Ba$_{3}$Ir$_{4}$Ge$_{16}$ and BaIr$_{2}$Ge$_{7}$ are new caged superconductors with $T_\mathrm{C}$ of 6.5, 6.1 and 3.2\,K, respectively, which have been recently {synthesized}~\cite{guo2013,ishida2014,duong2014,Zhao}. In contrast to their 3D counterpart, $\beta$-pyrochlore oxides and filled skutterudites, are {comprised} of 2D cage units {which are} formed by Rh(Ir)-Ge frameworks with a Ba atom {at the centre}. The higher value of the electron-phonon coupling constant ($\lambda_{e-ph}$) originates from the low-lying vibration modes derived from the {encapsulated} Ba cations and is responsible for the emergence of superconductivity in these compounds. Based on this, the possibility of rattling-related unconventional superconductivity has been widely investigated. In BaIr$_{2}$Ge$_{7}$ and Ba$_{3}$Ir$_{4}$Ge$_{16}$ there are two superconducting phases which evolve with pressure~\cite{pei2022pressure}. After suppressing the ambient-pressure superconducting (SC-I) state with initial increasing applied pressure (at 15 GPa T$_c$ is below 2 K for BaIr$_{2}$Ge$_{7}$), a new high pressure superconducting (SC-II) state emerges unexpectedly, with $T_C$ increased to a maximum of 4.4 K and 4.0 K for BaIr$_{2}$Ge$_{7}$ and Ba$_{3}$Ir$_{4}$Ge$_{16}$, respectively~\cite{pei2022pressure}. These compounds exhibit metal-like behaviour at ambient pressure. It is reported that the pressure-induced phonon softening caused by cage shrinkage is the cause of the second phase. 

The exploration of superconductivity in low-dimensional rhodium- and iridium-based filled caged compounds has ignited significant interest, prompting an in-depth investigation into their superconducting gap structure. This study delves into the microscopic aspects of superconductivity, focusing on Ba$_{3}$Ir$_{4}$Ge$_{16}$, employing a multi-faceted approach involving magnetization, heat capacity, and muon spin relaxation and rotation ($\mu$SR) measurements. This research marks the inaugural experimental exploration of superconductivity at the microscopic level, complemented and reinforced by concurrent theoretical calculations. Our comprehensive analysis of Ba$_{3}$Ir$_{4}$Ge$_{16}$ reveals it to be a conventional, nodeless $s$-wave superconductor characterized by an intermediate coupling strength. The superconducting order parameter and pairing mechanisms are elucidated through meticulous investigation, shedding light on the intriguing behavior of this compound in its superconducting state. The Fermi surface topography, a pivotal aspect of the study, exhibits complete compatibility with the isotropic $s$-wave symmetry of the superconducting ground state. Moreover, our findings point towards a relatively modest electron-rattler coupling. This research not only contributes to the growing body of knowledge regarding superconductivity in low-dimensional compounds but also establishes a foundation for further exploration in the realm of unconventional superconductors. The combination of experimental techniques and theoretical support enhances the robustness of our findings, providing a nuanced understanding of the superconducting properties of Ba$_{3}$Ir$_{4}$Ge$_{16}$ and offering valuable insights into the broader landscape of condensed matter physics.

\section{Methods}

\subsection{Experimental Details}

 A  polycrystalline specimen of Ba$_{3}$Ir$_{4}$Ge$_{16}$ was synthesized by precisely combining ultra-pure elements of Ba, Ir, and Ge chips in a stoichiometric ratio of 3:4:16. The synthesis process involved the arc melting of these constituent elements in a water-cooled copper crucible under an argon atmosphere. To enhance the sample's homogeneity, the resulting ingot underwent multiple cycles of flipping and remelting. Subsequently, the sample was annealed in evacuated quartz tubes at 1000$^{\circ}$C for a duration of 20 hours.

The phase purity of the synthesized sample was rigorously confirmed through X-ray diffraction (XRD) analysis. The XRD pattern, acquired at ambient temperature and pressure, was meticulously recorded using a Bruker D8 Advance X-ray diffractometer equipped with a copper rotating anode ($\lambda$ = 1.54058 \AA). The crystal structure was visually represented using \textsc{VESTA}~\cite{momma2008vesta}, a freely available software known for its versatility in crystallographic visualization.

Experimental investigations further delved into the physical properties of Ba$_{3}$Ir$_{4}$Ge$_{16}$ through a suite of techniques. Magnetization measurements were conducted using a vibrating sample magnetometer (MPMS, Quantum Design), providing insights into the material's magnetic behavior. Specific heat data was obtained using the traditional thermal relaxation method, with measurements performed down to an impressive temperature of 50 mK. The specific heat measurement process involved two distinct temperature regimes. Initially, data was collected from 290 K to 3 K utilizing liquid helium. Subsequently, measurements were extended down to 50 mK below 3 K, employing a combination of He$^3$ and He$^4$ within a dilution refrigerator.

This comprehensive experimental approach not only ensured the confirmation of the sample's structural integrity but also facilitated a thorough exploration of its magnetic and thermal properties. The synthesis and characterization process laid the foundation for a detailed understanding of Ba$_{3}$Ir$_{4}$Ge$_{16}$, contributing valuable insights to the broader field of condensed matter physics.

The $\mu$SR experiments detailed in this manuscript were conducted at the ISIS Pulsed Neutron and Muon Source in the United Kingdom, utilizing the MUSR spectrometer equipped with 64 detectors in both the forward (F) and backward (B) directions~\cite{Lee1999}. The sample, coated uniformly with GE varnish, was mounted on a highly pure (99.995\%) silver plate, ensuring a time-independent background. 100\% spin-polarized positive muons ($\mu^{+}$) were implanted into the sample, subsequently decaying into positrons after 2.2 $\mu$s.

The detectors recorded the number of positrons detected in both the forward and backward directions, denoted as $N_{\mathrm{F}}(t)$ and $N_{\mathrm{B}}(t)$, respectively. The asymmetry ($A(t)$) is expressed as $A(t)=\frac{N_{\mathrm{F}}(t)-\alpha N_{\mathrm{B}}(t)}{N_{\mathrm{F}}(t)+\alpha N_{\mathrm{B}}(t)}$, where $\alpha$ is a constant related to the MuSR instrument and estimated from a 20 G applied field in longitudinal mode.

TF-$\mu$SR measurements were conducted in the vortex state with a 300 Oe magnetic field applied perpendicular to the muon spin direction. Subsequently, ZF-$\mu$SR measurements were performed in longitudinal geometry. ZF-$\mu$SR serves as a sensitive local probe of small internal magnetic fields arising from the ordering of magnetic moments or moments that are randomly oriented and static or quasi-static during the muon's lifespan~\cite{sonier2000}. To minimize stray magnetic fields at the sample position, an active compensating system neutralized them to a level of approximately $\sim 0.001$ Oe.

The asymmetry spectra in both TF- and ZF-modes were collected over the temperature range of 1.3 K to 7.5 K using a Variox cryostat. All $\mu$SR data underwent analysis using the Muon Data Analysis (WiMDA) software, a freely available tool for Windows~\cite{Pratt2000}. These experimental procedures and analyses provide a robust foundation for investigating the magnetic properties of the material under scrutiny.

\subsection{Computational Methods}

We performed first-principles calculations to investigate the electronic structure using the Kohn-Sham scheme within the framework of Density Functional Theory (DFT)\cite{kohn1965}. Quantum Espresso was employed for these calculations, employing scalar-relativistic projector augmented wave pseudopotentials \cite{dal2014} for Ba, Ir, and Ge. Specifically, the Ba.pbe-spn-kjpaw\_psl.1.0.0, Ir.pbe-spn-kjpaw\_psl.1.0.0, and Ge.pbe-dn-kjpaw\_psl.1.0.0 pseudopotentials from the \texttt{PSlibrary} \cite{dal2014} were utilized. To account for Exchange and Correlation (XC) effects, we adopted the Generalized Gradient Approximation (GGA) within the Perdew-Burke-Ernzerhof (PBE) parametrization \cite{perdew1996}. The wave functions were computed with a kinetic energy cutoff of 70 Ry (1 Ry $\approx$ 13.6 eV), while a cutoff of 420 Ry was applied for charge density and potential. We employed a $16\times16\times16$ $k$-point sampling in the first Brillouin zone for $1.0\times10^{-8}$\, Ry self-consistent convergence, and a denser $24\times24\times24$ $k$-point sampling for detailed analysis including band structure, density of states, and Fermi surface. For self-consistent-field (SCF) and non-self-consistent field (NSCF) calculations, we implemented the Marzari-Vanderbilt smearing \cite{marzari1999} with a spreading of 0.005\, Ry for Brillouin-zone integration. All internal degrees of freedom were relaxed to ensure convergence of 10$^{-5}$\, Ry in total energy and 0.5\,mRy/$a_0$ ($a_0\approx0.529\,$Å) for forces acting on the nuclei. The lattice parameters were kept fixed at their experimental values throughout the calculations.

\begin{figure}[t]
\centering
\includegraphics[width=\linewidth]{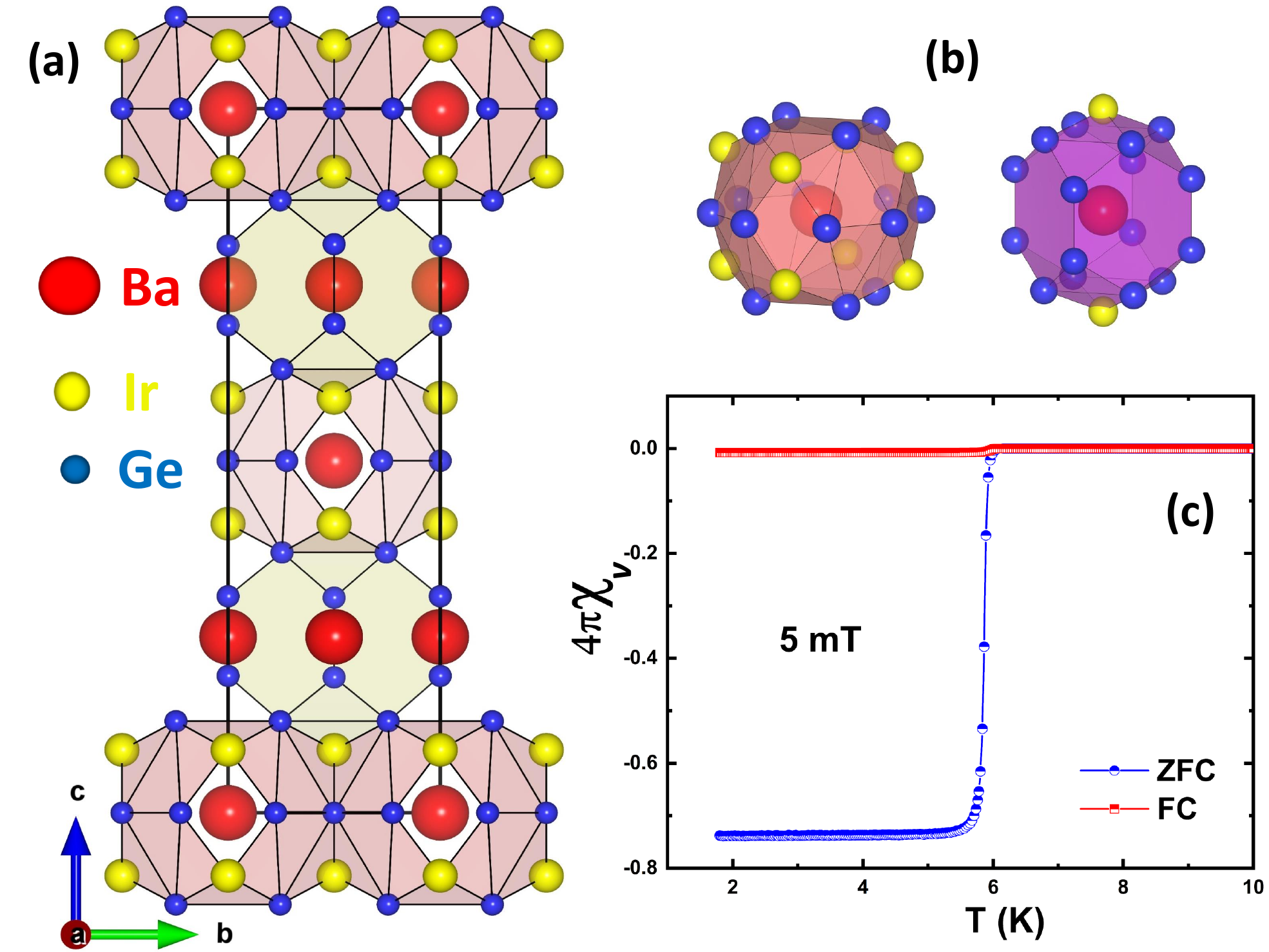}
\caption{(a-b) The crystal structure of Ba$_3$Ir$_4$Ge$_{16}$ is elucidated in this study, particularly focusing on the crystallographic $bc$ lattice plane. Detailed characterization of the arrangement and connectivity of atoms within the crystal lattice provides crucial insights into the material's structural properties. (c) To discern the superconducting behavior of Ba$_3$Ir$_4$Ge$_{16}$, we examines the temperature dependence of the zero-field-cooled (ZFC) and field-cooled (FC) superconducting volume fraction, denoted as $4\pi\chi_v$. The analysis is based on magnetization data collected at 5 mT, allowing for a comprehensive understanding of the material's superconducting properties under varying temperatures. This investigation contributes to the broader understanding of the superconducting characteristics of Ba$_3$Ir$_4$Ge$_{16}$.}
\label{mag}
\end{figure}

\par

\begin{figure}[t]
\centering
\includegraphics[height =\linewidth,width=0.8\linewidth]{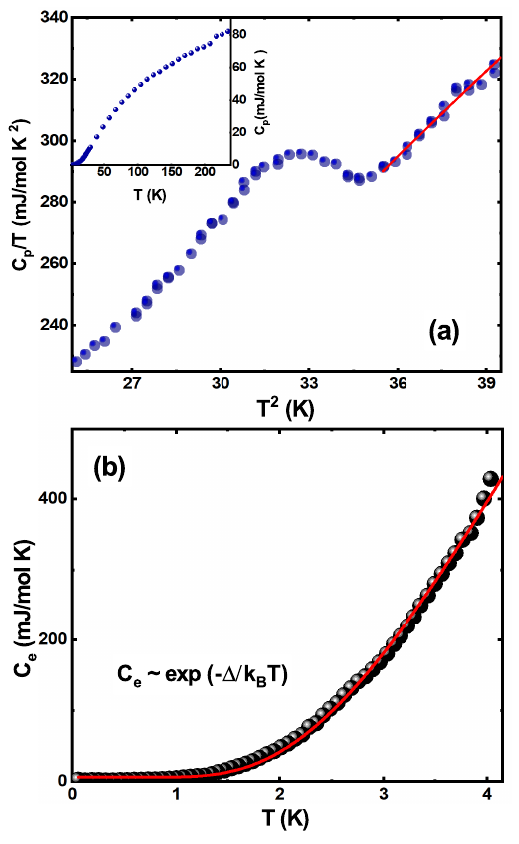}
\caption{ (a) The low-temperature heat capacity profile of Ba$_3$Ir$_4$Ge$_{16}$ is presented, offering a detailed examination of the material's thermal behavior in the specified temperature range. The inset provides a closer look at the variation of heat capacity across the temperature spectrum from 2 to 300 K, providing valuable insights into the specific thermal characteristics of the compound. (b) The electronic heat capacity of Ba$_3$Ir$_4$Ge$_{16}$ is  calculated by subtracting the phonon contribution, estimated from high-temperature data. This process allows for the isolation and in-depth analysis of the electronic component of heat capacity in the material. The red solid line in the corresponding graph represents a fit using the BCS gap model, offering a theoretical framework to interpret the observed electronic heat capacity behavior. This comprehensive analysis enhances our understanding of the thermal and electronic properties of Ba$_3$Ir$_4$Ge$_{16}$.}
\label{heat}
\end{figure}

\section{Results and discussion}

\begin{figure*}[t]
\centering
\includegraphics[width=0.9\linewidth]{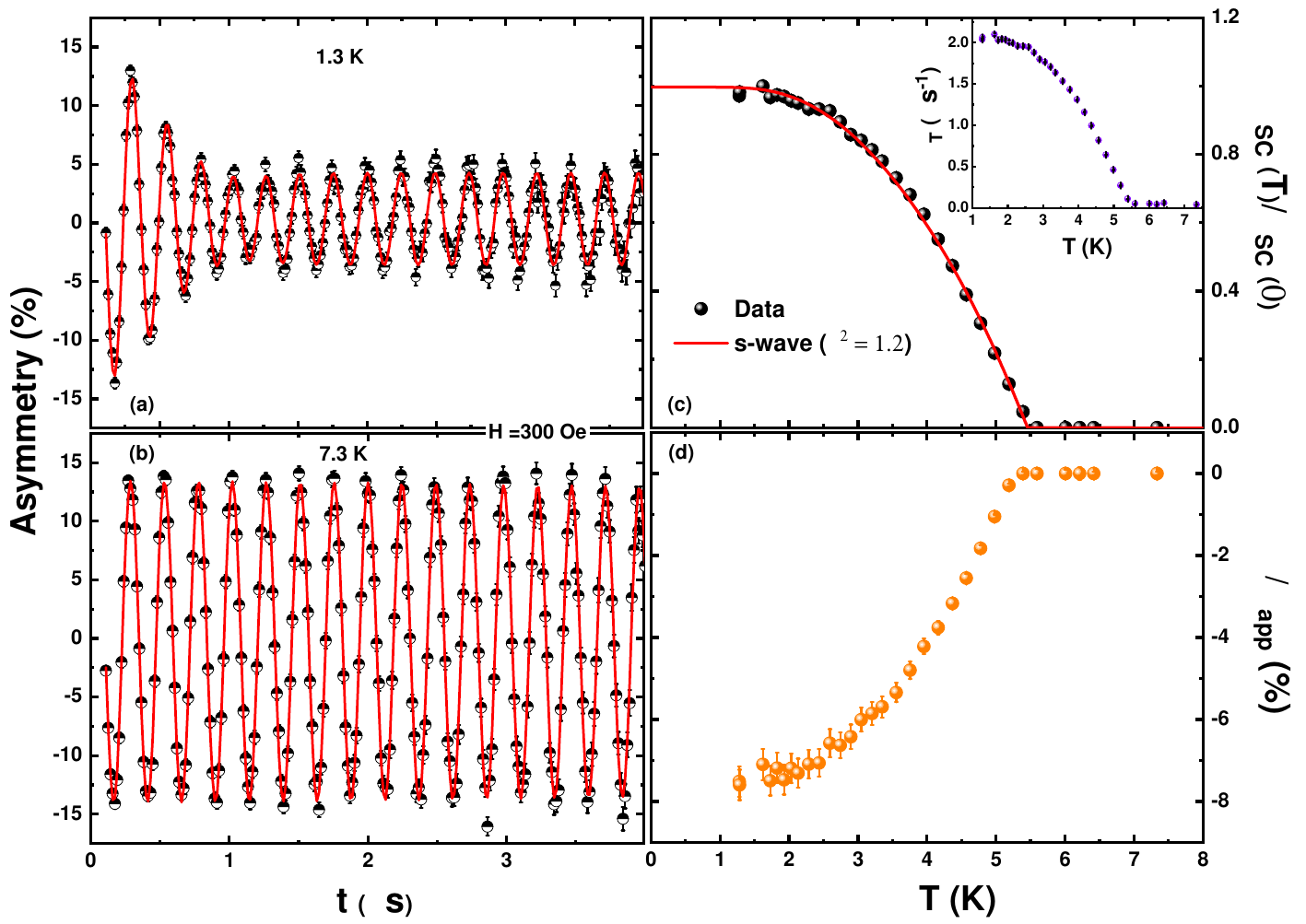}\hfil
\caption{(a) The temporal evolution of muon spin asymmetry spectra for Ba$_{3}$Ir$_{4}$Ge$_{16}$ is depicted in the zero-field-cooled (ZFC) mode. Panel (a) illustrates the spectra at two distinct temperatures, namely $T = 1.3$ K ($\leq T_\mathrm{C}$) and $T = 7.3$ K ($\geq T_\mathrm{C}$), the latter being within the vortex state. Employing Equation \ref{TF} detailed in the text, a solid red line represents the fitted data, offering a quantitative representation of the observed muon spin asymmetry spectra. (c) The temperature dependence of the superfluid density $\sigma_\mathrm{sc}(T)/\sigma_\mathrm{sc}(0)$ is presented. A fitting approach utilizing an isotropic $s$-wave model, as described in the text, is indicated by the solid red line. This analysis provides valuable insights into the behavior of the superfluid density as a function of temperature. The inset showcases the temperature dependence of the total superconducting depolarization rate $\sigma_\mathrm{T}$. (d) The temperature evolution of the shift of the internal magnetic field in the vortex state is illustrated. This observation contributes to our understanding of the dynamic changes in the internal magnetic environment of Ba$_{3}$Ir$_{4}$Ge$_{16}$ under varying temperatures, providing critical information about the material's superconducting properties.}
\label{musrdata}
\end{figure*}

\subsection{Crystal Structure \& Physical Properties}

The crystal structure of Ba$_3$Ir$_4$Ge$_{16}$, depicted in Fig. 1(a-b), is characterized by a tetragonal symmetry (space group $I4/mmm$). The lattice parameters $a$ and $c$ are determined to be $6.5387$~\AA ~and $22.2834$~\AA, respectively, consistent with earlier reports~\cite{ishida2014}. Upon substituting $Ir$ for $Rh$, the $a$-axis expands while the $c$-axis contracts, resulting in an approximately invariant cell volume when compared to Ba$_3$Rh$_4$Ge$_{16}$. Notably, in contrast to 3D cage-forming frameworks, this compound exhibits a 2D network structure interwoven with layers of Ir-Ge cages. The encapsulation of barium atoms in [Ir$_8$Ge$_{16}]^{2-}$ cages, connected by [Ir$_2$Ge$_{16}$] cages, defines the distinctive crystal arrangement. The crystal structure adopts an interlayer face-sharing connection, resembling a stack of layers alternately composed of Ba\texttt{@}[Ir$_2$Ge$_{16}$] and Ba\texttt{@}[Ir$_8$Ge$_{16}$]. This layered configuration provides additional space for Ba cations, facilitating an increase in anharmonic vibration components. Consequently, the rattling frequency of Ba cations is reduced in this structural arrangement.

The low-temperature DC susceptibility, illustrated in Fig. \ref{mag}(c), showcases the temperature dependence in both zero-field cool (ZFC) and field cool (FC) modes, confirming the manifestation of bulk superconductivity in the material. Considering the spherical powder sample used for magnetization measurements, a demagnetization factor of $D = 1/3$ is assumed~\cite{das2021probing}. The estimated superconducting volume fraction of 74\% aligns well with the findings from transverse-field muon spin rotation (TF-$\mu$SR) data. In the ZFC domain, when the external magnetic field is activated while the sample is superconducting, it behaves as a perfect diamagnet. In the FC regime, it shields the change in magnetic field rather than the magnetic field itself. Above 6 K, no magnetic phase transition is discerned.

Beyond the transition temperature, the resistivity exhibits a square dependence on temperature, indicative of significant electron-electron scattering~\cite{pei2022pressure}. The upper critical field and Ginzburg-Landau coherence length are estimated to be 18.2 kOe and 13.4 nm, respectively~\cite{guo2013}. These observations contribute valuable insights into the superconducting properties and electronic behavior of the material under investigation.

In conjunction with magnetization measurements, heat capacity data has been acquired at zero field, as illustrated in Fig.~\ref{heat}(a). A discernible jump at around 5.7 K in $C_p(T)$ unequivocally signals a superconducting transition, consistent with the findings from $\chi(T)$. The normal state-specific heat data is subjected to fitting using $C_{p} = \gamma T + \beta T^3 + \eta T^5$, where $\gamma$ denotes the Sommerfeld coefficient, $\beta T^3$ is attributed to the Debye model, and an additional $\eta T^5$ term is introduced to account for the anomalous dispersion of the phonon spectrum within the intricate crystal structure. The fitting yields $\gamma = 80.1\pm0.2$ mJ/(mol$^{-1}$K$^{-4}$) and $\beta = 3.3\pm0.12$ mJ mol$^{-1}$K$^{-2}$. Utilizing the Debye relation, which incorporates the gas constant $R = 8.314$ J mol$^{-1}$K$^{-1}$ and the number of atoms per formula unit $n = 23$ in Ba$_{3}$Ir$_{4}$Ge$_{16}$, the Debye temperature ($\Theta_{D}$) is determined to be $238\pm2$ K. This value, although slightly higher, is in good agreement with the corresponding value for Ba$_3$Rh$_4$Ge$_{16}$ ($\Theta_{D}$ = 221 K)~\cite{zhao2021superconductivity}. The normal state data exhibits linearity at higher temperatures, and for clarity, a limited range is presented for visual inspection.

Upon subtracting the phonon contribution, the electronic specific heat $C_{e}$ (T) is extracted and depicted in Fig.~\ref{heat}(b). The temperature dependence of electronic heat capacity serves as a crucial indicator of the superconducting gap structure. A conventional exponential temperature dependence fits the data most effectively (\textcolor{black}{$C_{e}\sim \exp{\Delta}/{k_{B}T}$}), revealing a superconducting gap $\Delta$(0) of 0.88 meV, closely aligned with the value of 0.96 meV obtained from the analysis of TF-$\mu$SR data, as discussed subsequently. The ratio of the superconducting gap to $T_\mathrm{C}$ is computed to be 3.70 based on the heat capacity data, providing additional insights into the superconducting behavior of the material.

\subsection{Superconducting Gap Structure}

The investigation of the superconducting gap structure in Ba$_{3}$Ir$_{4}$Ge$_{16}$ was conducted through TF-$\mu$SR measurements in the vortex state, offering valuable insights. Figs.~\ref{musrdata}(a)-(b) present the time evolution of TF$\mu$SR asymmetry spectra both below and above the critical temperature ($T_\mathrm{C}$). In the vortex state, the field distribution of the flux line lattice exhibits inherent inhomogeneity, leading to observed damping effects in the asymmetry spectra below $T_\mathrm{C}$. To precisely characterize this behavior, the data were subjected to fitting procedures employing an oscillating Gaussian decay function representing the muon spin dynamics. Additionally, an oscillating background attributed to the silver sample holder was considered in the model~\cite{bhattacharyya2018, adroja2017, bhattacharyya2019}. The fitting function utilized is expressed as:
\begin{equation}
G_\mathrm{TF}(t) = A_\mathrm{sc}\cos(\omega_\mathrm{1}t+\phi)\exp\left(-\frac{\sigma_{T}^{2}t^{2}}{2}\right)+A_\mathrm{bg}\cos(\omega_\mathrm{2}t+\phi),
\label{TF}
\end{equation} 
where $A_{\mathrm{sc}}$ represents the asymmetry (70.7\%) of the muon signal originating from the sample, and $A_{\mathrm{bg}}$ denotes the asymmetry (29.3\%) of the signal associated with the sample that interacts with the holder. This formulation captures the complex interplay of oscillatory components and decay processes, providing a comprehensive description of the observed TF-$\mu$SR asymmetry spectra.

During the fitting procedure, $A_{\mathrm{bg}}$ was held constant at a fixed value of 0.293, determined at the lowest available temperature of 1.3 K. This choice was informed by several factors, including the powder nature of the sample (with a packing fraction below 100\%), a beam size slightly exceeding the sample dimensions, and the inherent beam divergence. Such asymmetry values for the sample (70.7\%) and the holder (29.3\%) are well-founded, aligning with observations from various $\mu$SR experiments~\cite{adroja2021pairing, svanidze2015non}. The rationale for these asymmetry values is corroborated by the asymmetry profile depicted in Fig. 3 (a, b). In proximity to $t$ = 0, a 15\% asymmetry is discerned, consistent with previous analyses~\cite{singh2014detection, biswas2012comparative, bhattacharyya2019investigation}. Furthermore, at larger time scales, a non-decaying asymmetry of approximately 5\% is observed. While acknowledging potential contributions from decaying components to this 5\% asymmetry, a favorable agreement is noted with the values of $A_1$ and $A_2$ estimated through the fitting of Eq.~\ref{TF}.

The muon precession frequencies, denoted as $\omega_{\mathrm{1}}$ and $\omega_{\mathrm{2}}$, correspond to the sample and sample holder, respectively, and are intimately linked to the internal field distribution. The temperature-dependent shift of the internal magnetic field in the vortex state is delineated in Fig.~\ref{musrdata}(d). The initial phase, $\phi$, assumed to be identical for both the sample and background, was held constant at its lowest temperature value throughout the analysis. The total depolarization rate, $\sigma_{T}$, is composed of two components: (i) the superconducting contribution, $\sigma_\mathrm{sc}$, and (ii) the normal state contribution, $\sigma_{\mathrm{n}} = 0.0517 \mu s^{-1}$, which remains temperature-independent. The superconducting contribution, $\sigma_\mathrm{sc}$, is calculated by subtracting the normal state contribution using the relation $\sigma_\mathrm{sc}=\sqrt{\sigma_\mathrm{T}^{2}-\sigma_\mathrm{n}^2}$. This comprehensive approach ensures a detailed and accurate characterization of the TF-$\mu$SR measurements and provides valuable insights into the superconducting gap structure of Ba$_{3}$Ir$_{4}$Ge$_{16}$.

\par

To unravel the intricacies of the superconducting gap structure, we scrutinize the temperature dependence of the normalized $\frac{\sigma_\mathrm{sc}(T)}{\sigma_\mathrm{sc}(0)}$, a parameter intimately connected to the superfluid density. Our modeling approach, guided by established methodologies~\cite{prozorov2006}, involves the expression:

\begin{eqnarray}
\frac{\sigma_\mathrm{sc}(T)}{\sigma_\mathrm{sc}(0)} &= \frac{\lambda^{-2}(T,\Delta_\mathrm{0,i})}{\lambda^{-2}(0,\Delta_\mathrm{0,i})}\\ \nonumber
 &= 1 + \frac{1}{\pi}\int_{0}^{2\pi}\int_\mathrm{\Delta(T)}^{\infty}(\frac{\delta f}{\delta E}) \times \frac{EdEd\phi}{\sqrt{E^{2}-\Delta(T,\phi})^2},
\end{eqnarray}

where the Fermi function $f$ is defined as $f= [1+\exp(E/k_\mathrm{B}T)]^{-1}$. The superconducting gap, denoted as $\Delta$, is a function of temperature ($T$) and polar angle for anisotropy ($\phi$) and is expressed as $\Delta(T,\phi) = \Delta_\mathrm{0}\delta(T/T_\mathrm{C})\mathrm{g}(\phi)$. Here, $\Delta_\mathrm{0}$ signifies the gap value at absolute zero temperature, and $\delta(T/T_\mathrm{C}) = \tanh[1.82[1.018(T_\mathrm{C}/T-1)]^{0.51}]$. The angular dependence of the superconducting gap function, represented by $\mathrm{g}(\phi)$, is equal to 1 for an isotropic $s$-wave model~\cite{annett1990}. Our analysis, employing a single isotropic $s$-wave gap with $\Delta_\mathrm{0}$ set at 0.96\,meV, yields a gap to $T_\mathrm{C}$ ratio of 2$\Delta/k_\mathrm{B}T_\mathrm{C} = 4.04$. This value surpasses the theoretical BCS limit for a weak-coupling superconductor (3.53), placing the compound within the category of moderate-coupling superconductors~\cite{BCS, lin2003bcs, mu2007possible, sun2004magnetic}. It is noteworthy that data acquisition below 1.3~K was omitted, as the $\mu$SR data down to this temperature aligns with heat capacity and an $s$-wave model. The absence of novel physics in the millikelvin range underscores the completeness of our dataset in providing a comprehensive understanding of the superconducting behavior in Ba$_{3}$Ir$_{4}$Ge$_{16}$.

\begin{figure}[h]
\centering
\includegraphics[width=\linewidth]{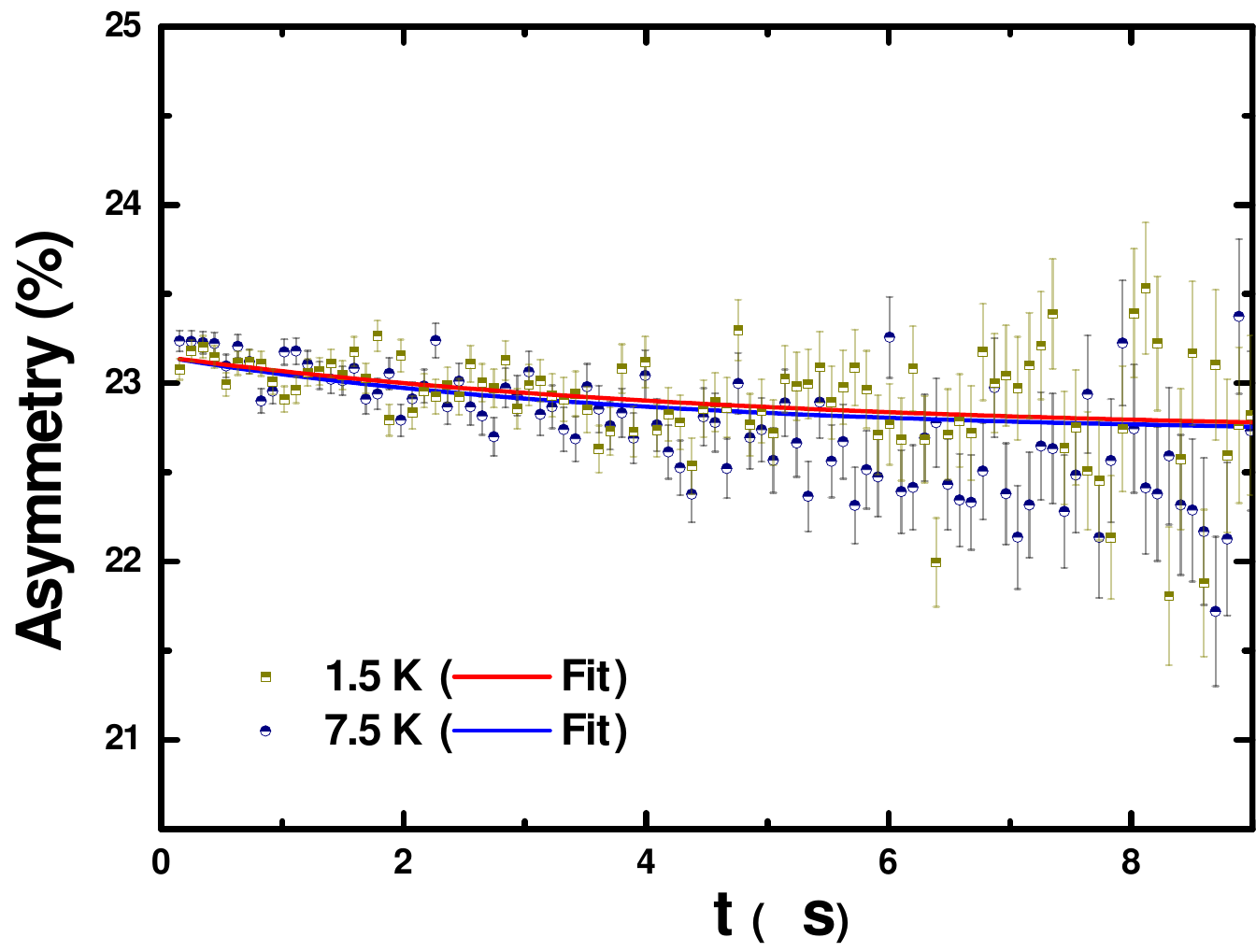}
\caption{ The time evolution of the zero-field (ZF) $\mu$SR asymmetry spectra for Ba$_{3}$Ir$_{4}$Ge$_{16}$ is portrayed at two distinct temperatures, specifically at 1.5 K (indicated by dark yellow squares) and 7.3 K (represented by navy circles). The solid red line overlaid on the data signifies the least square fit, performed using Equation as detailed in the text. Notably, the fit to the data exhibits a striking similarity for both temperatures, underscoring the robustness of the model in capturing the intricate features of the ZF-$\mu$SR asymmetry spectra across the specified temperature range.}
\label{ZF}
\end{figure}

\subsection{Superconducting Parameters}

The relationship between the muon spin depolarization rate below $T_\mathrm{C}$ and the London penetration depth ($\lambda_L$) is expressed by the equation $\frac{\sigma_\mathrm{sc}^2(T)}{\gamma_\mathrm{\mu}^2}=\frac{0.00371 \times \phi_\mathrm{0}^{2}}{\lambda^4_L(T)}$~\cite{amato1997,chia2004}. Here, $\phi_{\mathrm{0}}$ represents the flux quantum number, taking the value of 2.07 $\times$10$^{-15}$Tm$^{2}$, and $\gamma_{\mathrm{\mu}}$ is the muon gyromagnetic ratio, with $\gamma_\mathrm{{\mu}}/2\pi$ equal to 135.5\,MHzT$^{-1}$. Within the confines of the s-wave model, the London penetration depth at zero temperature, $\lambda_L$(0), is approximated to be 190 nm. London's theory provides an avenue to estimate additional phenomenological parameters characterizing the superconducting state. Specifically, the relation $\lambda_{\mathrm{L}}^2 = \frac{m^{*}c^{2}}{4\pi n_\mathrm{s}e^{2}}$~\cite{sonier2000} unveils insights into the system. Here, $m^{*} = (1+\lambda_\mathrm{e-ph})m_\mathrm{e}$ denotes the effective mass, with $\lambda_\mathrm{e-ph}$ representing the electron-phonon coupling parameter, $m_\mathrm{e}$ being the electron mass, and $n_\mathrm{s}$ serving as the superconducting carrier density. By unraveling these relationships, we gain a comprehensive understanding of the interplay between muon spin dynamics and the underlying superconducting state, shedding light on the intricate physical properties of the Ba$_{3}$Ir$_{4}$Ge$_{16}$ compound. The electron-phonon coupling parameter, denoted as $\lambda_{\mathrm{e-ph}}$, is a crucial descriptor quantifying the interaction between electrons and phonons in a superconducting material. It can be calculated using McMillan's formula~\cite{mcmillan1968,ferreira2018,bhattacharyya2019_2}, which is expressed as:

\begin{equation}
\lambda_\mathrm{e-ph} = \frac{1.04+\mu^{*}\ln(\Theta_\mathrm{D}/1.45T_\mathrm{C})}{(1-0.62\mu^{*})\ln(\Theta_\mathrm{D}/1.45T_\mathrm{C})-1.04}.
\end{equation}

Here, $\mu^{*}$ represents the repulsive screened Coulomb parameter,\textcolor{black}{~ranging from $\mu^{*}$= 0.10-0.16, for which here we have use an average value of $\mu^{*}$ = 0.13~\cite{Errea2015,Talantsev2020,mcmillan1968}}. The resulting electron-phonon coupling parameter for Ba$_{3}$Ir$_{4}$Ge$_{16}$ is $\lambda_{\mathrm{e-ph}} = 0.71\pm 0.01$. This value aligns well with prior reports~\cite{guo2013superconductivity} and closely resembles the electron-phonon coupling parameter of Ba$_3$Rh$_4$Ge$_{16}$ (0.8). As Ba$_{3}$Ir$_{4}$Ge$_{16}$ falls into the category of type-II superconductors, where almost all normal state carriers ($n_\mathrm{e}$) contribute to superconductivity ($n_\mathrm{s} \approx n_\mathrm{e}$), we can derive the superconducting carrier density $n_\mathrm{s}$ and the effective-mass enhancement $m^{*}$. The calculated values are $n_\mathrm{s} = 1.33\pm0.02 \times 10^{27}$\,carriers\,m$^{-3}$ and $m^{*} = 1.71\pm 0.01~m_\mathrm{e}$, respectively. These parameters offer valuable insights into the nature of superconductivity in Ba$_{3}$Ir$_{4}$Ge$_{16}$, providing a basis for further understanding the underlying physics of this intriguing material.

\begin{figure}[t]
    \centering
	\includegraphics[height =\linewidth,width=\linewidth]{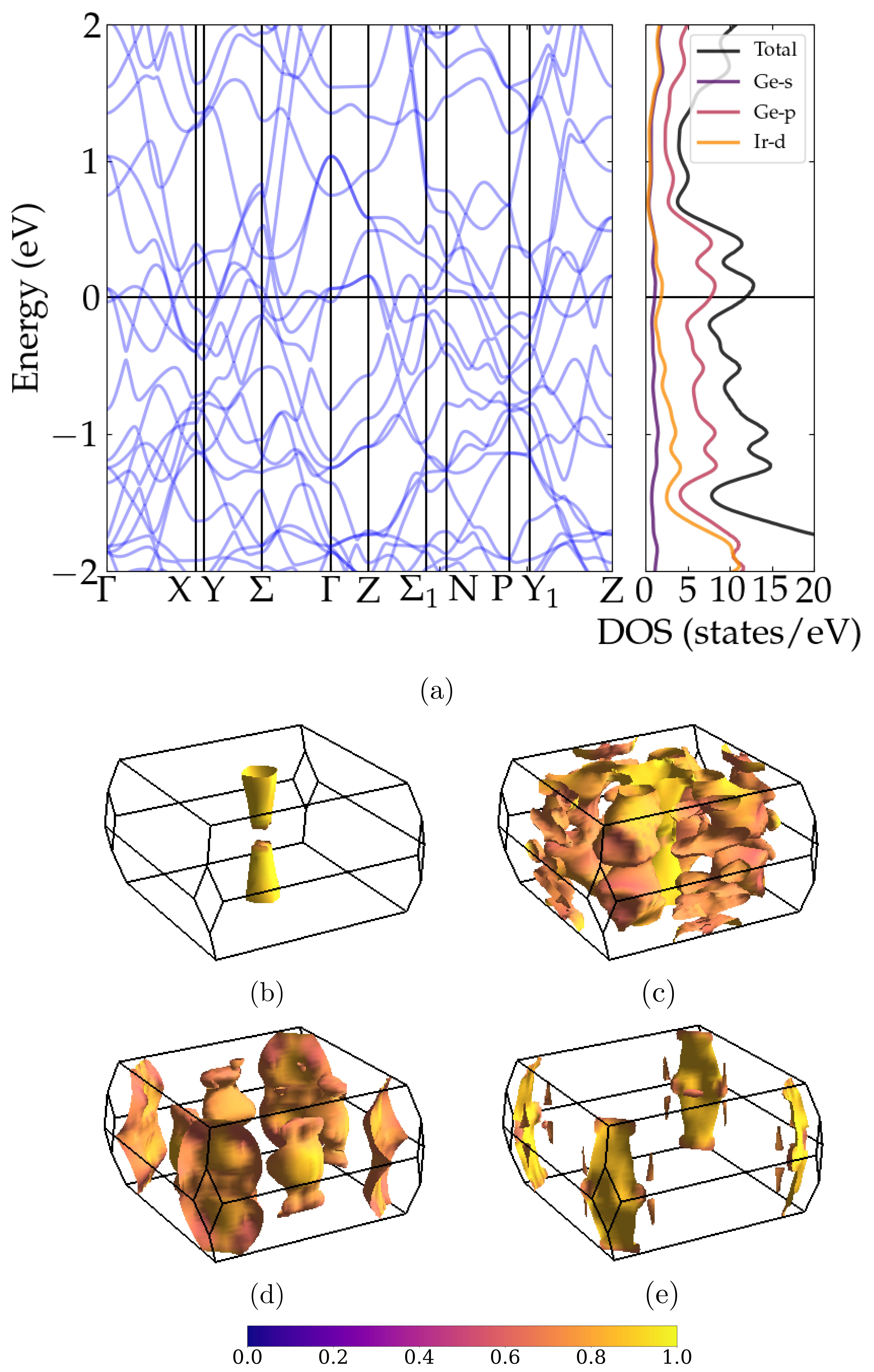}
	\caption{ (a) The electronic band structure and partial density of states without spin-orbit coupling (SOC) for Ba$_3$Ir$_4$Ge$_{16}$ are presented. This comprehensive analysis offers valuable insights into the material's electronic properties, highlighting the distribution of electronic states across the energy spectrum in the absence of SOC. (b)-(e) Fermi surface projections onto Ge-$p$ derived states are depicted in these panels. Each panel provides a distinct perspective on the Fermi surface, emphasizing the contributions of Ge-$p$ derived states. The nuanced variations in the Fermi surface across different projections contribute to a comprehensive understanding of the material's electronic structure and its dependence on specific electronic states derived from Ge-$p$ orbitals.}
	\label{fig:bands}
\end{figure}

\subsection{Zero field $\mu$SR}

Zero-field transverse-field muon spin rotation (ZF-$\mu$SR) serves as a powerful tool for investigating the possible presence of a spontaneous magnetic field in the superconducting state of Ba$_{3}$Ir$_{4}$Ge$_{16}$. The comparison of zero-field time-dependent asymmetry spectra above and below $T_\mathrm{C}$ (at $T = 1.5$ K and $7.3$ K) is presented in Fig.~\ref{ZF}. Strikingly, the obtained spectra are virtually indistinguishable in both cases. The absence of muon spin precession provides clear evidence against the existence of internal magnetic fields, a characteristic feature observed in magnetically structured compounds. The ZF-$\mu$SR data are effectively described by a Lorentzian function accompanied by a constant temperature-independent background term ($A_\mathrm{bg}$),

\begin{equation}
G_\mathrm{ZF}(t) = A_\mathrm{SC} \exp{(-\lambda t)} + A_\mathrm{bg},
\label{ZFfunction}
\end{equation}

where $A_\mathrm{SC}$ represents the zero-field asymmetry of the sample, and $\lambda$ denotes the muon spin relaxation rate attributable to randomly oriented nuclear moments. This analysis aids in providing a robust characterization of the magnetic behavior in the superconducting state of Ba$_{3}$Ir$_{4}$Ge$_{16}$.

After fixing the background ($A_{bg}$) at 1.5 K we estimated the $A_\mathrm{SC}$ = 1.9 and 2.2 for 7.3 K and 1.5 K respectively which is obvious due to the emergence of the superconducting state below 5.7 K. The {fit} to the ZF$-\mu$SR asymmetry data using Eq.~\ref{ZFfunction} is shown by the solid red line in Fig.~\ref{ZF}. It was found $\lambda = 0.317 ~\mu \mathrm{s}^{-1}$ at  1.5\,K and  $\lambda = 0.298 ~\mu \mathrm{s}^{-1}$ at 7.3\,K. There is no noticeable change between the relaxation rates at 7.3\, K ($\geq T_c$) and 1.5\, K ($\leq T_c$). These values of $\lambda$ at $T\leq T_{c}$ and $ \geq T_{c}$ agree within the expected error, indicating that time reversal symmetry is preserved in the superconducting state of Ba$_{3}$Ir$_{4}$Ge$_{16}$. {Bulk nature of superconductivity of the sample is also supported through the analyses of our TF-$\mu$SR (also ZF-$\mu$SR) data where we need to use only one superconducting component in addition to a silver holder background (A$_\mathrm{bg}$), which again support the bulk nature of superconductivity in Ba$_{3}$Ir$_{4}$Ge$_{16}$. If we had two phases, superconducting and non-superconducting below $T_\mathrm{C}$ then we need to account for the non-superconducting part by adding an extra component in Eq.~\ref{TF}, which was not the case in our analysis}.

\begin{table}
\caption{Comparison of key parameters between Ba$_{3}$Ir$_{4}$Ge$_{16}$ and Ba$_{3}$Rh$_{4}$Ge$_{16}$. The values include lattice parameters ($a$, $b$, $c$), critical temperature ($T_\mathrm{C}$), upper critical field ($H_\mathrm{c2}$), electronic specific heat coefficient ($\gamma$), Debye temperature ($\Theta_\mathrm{D}$), London penetration depth at zero temperature ($\lambda_\mathrm{L}$(0)), electron-phonon coupling parameter ($\lambda_\mathrm{e-ph}$), superfluid density ($n_\mathrm{s}$), and the ratio of the superconducting energy gap to the thermal energy at $T_\mathrm{C}$ (2$\Delta$/$k_\mathrm{B} T_\mathrm{C}$).}

\centering
\begin{tabular}{lcc}
\hline
Parameters & Ba$_{3}$Ir$_{4}$Ge$_{16}$ &  Ba$_{3}$Rh$_{4}$Ge$_{16}$ \\  \hline
$a$(\AA)     & 6.5387 & 6.5640\\
$b$(\AA)     & 6.5387 & 6.5640 \\
$c$(\AA)     & 22.2834 & 22.037\\
$T_\mathrm{C}$     & 5.7 & 7.0 \\
$H_\mathrm{c2}(T)$ & 2.1  & 2.5\\
$\gamma$(0) (mJ/mol K$^{2}$) &  21 &21.1  \\
$\Theta_\mathrm{D}$ & 238 & 221  \\
$\zeta_{0}$(nm) &  --   & 18.1\\
$\lambda_\mathrm{L}$(0)(nm) & 190 & 142 \\
$\lambda_\mathrm{e-ph}$ & 0.71& 0.80 \\
$n_\mathrm{s}$(carriers/m$^{3})$ & 1.33 $\times$ 10$^{27}$ & 2.3 $\times$ 10$^{27}$ \\
2$\Delta$/$k_\mathrm{B} T_\mathrm{C}$ & 4.04 &  3.52\\  \hline \hline

\end{tabular}
\label{tab:1}
\end{table}

\subsection{First principles calculations}

\noindent Figures \ref{fig:bands}(a)-(e) illustrate the electronic band structure, the partial density of states (DOS), and the Fermi surface projected onto Ge-$p$ orbitals of Ba$_3$Ir$_4$Ge$_{16}$ without considering the effect of spin-orbit coupling. One can observe a large density of states at the Fermi level of 11.8\,eV$^{-1}$ per unit cell due to four bands crossing the Fermi energy. Approximately 65\,\% of the carriers at the Fermi surface come from the Ge-$p$ manifold, 15\,\% from the Ir-$d$, and 9\,\% from Ge-$s$ orbitals. Such a high density of states favors spontaneous symmetry breaking as superconductivity. There is a total of 0.653\, states/eV of Ba-derived orbitals at the Fermi level, representing approximately $5.5\,\%$ of the free, low-energy states available. These values agree with previous first-principles calculations \cite{guo2013,ishida2014}. The electronic configuration suggests that the structure is composed of high-conductivity Ir--Ge layers and low-conductivity Ba--Ge layers, as first suggested by Ishida et\,al.~\cite{ishida2014}. The small but finite amount of Ba-derived orbitals composing the Fermi surface indicates that the guest atom is not completely ionized to form strong rattling modes, thus resulting in a negligible electron-rattler coupling. 

Additionally, the topography of the Fermi surface also indicates that unconventional pairing is very unlikely. Spin-fluctuation-induced unconventional superconductivity, for instance, is associated with a peak in the real part of the bare Lindhard susceptibility \cite{berk1966,fay1980,moriya2003}
\begin{equation}
\chi_0(q,w) = \sum_{k,m,n}\mid M_{k,k+q}^{m,n}\mid^2 \frac{f(\epsilon_k^m)-f(\epsilon_{k+q}^n)}{\epsilon_k^m - \epsilon_{k+q}^n - \omega - t\delta}
\end{equation}
where $\epsilon_k^m$ is the eigenvalue of band $m$ at wave vector $k$, $f$ is the Fermi distribution, and $M$ is the matrix element, which is commonly set to unity. Peaks in susceptibility reflect the nesting degree along separate sheets. However, it is clear looking at the Ba$_3$Ir$_4$Ge$_{16}$ Fermi surface in Figures \ref{fig:bands}(b)-(e) that there are no possible nesting regions that can be accounted for magnetic instabilities. As a consequence, the three-dimensional character of the Fermi surface excludes inter-band nesting-driven unconventional pairing mechanisms, as compared to their two-dimensional counterparts.

Moreover, in the realm of multi-gap superconductors, it is common to observe distinct orbital characteristics on different sheets of the Fermi surface. This results in an average electron-phonon scattering pattern that connects various points on the Fermi surface, often disjointed with respect to the band index \cite{floris2007, bersier2009, flores2015, heil2017, bhattacharyya2020, zhao2020, de2021, correa2021}. However, the Fermi surface of Ba$_3$Ir$_4$Ge$_{16}$, despite its multiband nature, stands out due to its homogeneous distribution of Ge-$p$ and Ir-$d$ hybridization across different sheets. This distinctive feature positions Ba$_3$Ir$_4$Ge$_{16}$ as a promising candidate for hosting a single superconducting gap characterized by conventional symmetry, primarily attributed to a substantial contribution from inter-band coupling \cite{kawamura2017, pascut2019, panda2019, bhattacharyya2021}. Consequently, the coherent mechanism of a single-band electron-phonon $s$-wave pairing emerges as a plausible explanation for the observed superconductivity in Ba$_3$Ir$_4$Ge$_{16}$, aligning well with our experimental findings.

\section{Summary}
 
In this study, we present a comprehensive analysis of the layered caged compound Ba$_3$Ir$_4$Ge$_{16}$, classifying it as an electron-phonon superconductor within a moderate coupling limit characterized by isotropic single-gap $s$-wave pairing symmetry. The temperature-dependent normalized superfluid density, derived from time-resolved transverse-field muon spin rotation (TF-$\mu$SR) spectra, reveals a spherically symmetric superconducting gap with $2\Delta(0)/k_BT_c = 4.04$. This observation aligns with a scenario of moderate phonon-mediated pairing instability. From the experimentally determined critical temperature ($T_\mathrm{C}$) of 5.7 K, as obtained from zero-field-cooled (ZFC) and field-cooled (FC) magnetization measurements, and the reported Debye temperature, we derive a mass enhancement parameter of 0.71. This value indicates a slightly smaller interaction compared to Ba$_3$Rh$_4$Ge$_{16}$. Zero-field (ZF)-$\mu$SR time spectra reveal the absence of a spontaneous magnetic field below $T_c$, confirming the preservation of time-reversal symmetry.

First-principles electronic-structure calculations corroborate previous findings, identifying four bands crossing the Fermi level. Ge-derived states, predominantly featured, exhibit a robust hybridization with Ir-$d$ orbitals, uniformly distributed across different sheets of the Fermi surface. The homogeneous disconnected multiband nature of the Fermi surface indicates substantial inter-band coupling, favoring the emergence of single-gap superconductivity. Notably, the presence of Ba-derived states from crown-shaped Ba-Ge rings at the Fermi level suggests a negligible electron-rattler interaction, excluding unconventional pairing symmetries. The absence of nesting spin- and charge instabilities further rules out unconventional superconductivity mediated by magnetic fluctuations.

Our findings underscore Ba$_{3}$Ir$_{4}$Ge$_{16}$ as a structurally and electronically distinct member of the superconducting polyhedral caged materials' family. The unique quasi 2D networks, composed of crown-shaped Ge rings that cage Ba atoms at the center, open new avenues for exploring the interplay among crystalline symmetries, low-dimensional structure units, anharmonic modes, and superconducting pairing mechanisms.
 
\section*{Acknowledgements}

AB expresses gratitude to the Science and Engineering Research Board for the CRG Research Grant (CRG/2020/000698 \& CRG/2022/008528) and CRS Project Proposal at UGC-DAE CSR (CRS/2021-22/03/549). AB thanks the Department of Science and Technology, India (SR/NM/Z-07/2015) for providing access to the experimental facility and financial support for conducting the experiment. Additionally, Jawaharlal Nehru Centre for Advanced Scientific Research (JNCASR) is acknowledged for project management.

DTA acknowledges the Royal Society of London for the UK-China Newton funding - EPSRC UK (Grant Ref: EP/W00562X/1) and the Japan Society for the Promotion of Science for an invitation fellowship. AKJ expresses gratitude to DST India for the Inspire Fellowship (IF190793). RT acknowledges the Indian Nanomission for post-doctoral funding. YQ extends thanks to the National Key R \& D Program of China (Grant No. 2018YFA0704300) and the National Natural Science Foundation of China (Grant No. U1932217 and 11974246).

This research was partly financed by the Coordenação de Aperfeiçoamento de Pessoal de Nível Superior – Brasil (CAPES) – Finance Code 001. The authors are grateful for the financial support provided by the São Paulo Research Foundation (FAPESP) under Grants 2019/05005-7 and 2020/08258-0. The study utilized high-performance computing resources made available by the Superintendência de Tecnologia da Informação (STI), Universidade de São Paulo. The authors also acknowledge the National Laboratory for Scientific Computing (LNCC/MCTI, Brazil) for providing HPC resources of the SDumont supercomputer, which have significantly contributed to the research reported in this paper.

\bibliographystyle{elsarticle-num}
\bibliography{references} 

\begin{thebibliography}{10}
\expandafter\ifx\csname url\endcsname\relax
  \def\url#1{\texttt{#1}}\fi
\expandafter\ifx\csname urlprefix\endcsname\relax\def\urlprefix{URL }\fi
\expandafter\ifx\csname href\endcsname\relax
  \def\href#1#2{#2} \def\path#1{#1}\fi

\bibitem{lee2006}
P.~A. Lee, N.~Nagaosa, X.~G. Wen, Reviews of modern physics 78 (2006) 17.

\bibitem{stewart2011}
G.~Stewart, Reviews of modern physics 83 (2011) 1589.

\bibitem{pei2022pressure}
C.~Pei, T.~Ying, Y.~Zhao, L.~Gao, W.~Cao, C.~Li, H.~Hosono, Y.~Qi, Matter and radiation at extremes 7 (2022) 038404.

\bibitem{kelty1991}
S.~P. Kelty, C.~C. Chen, C.~M. Lieber, Nature 352 (1991) 223--225.

\bibitem{ellis2007caged}
G.~C. Ellis-Davies, Nature methods 4 (2007) 619--628.

\bibitem{rosseinsky1991}
M.~J. Rosseinsky, A.~Ramirez, S.~Glarum, D.~Murphy, R.~Haddon, A.~Hebard, T.~Palstra, A.~Kortan, S.~Zahurak, A.~Makhija, Physical review letters 66 (1991) 2830.

\bibitem{hebard1991}
A.~Hebard, M.~Rosseinky, R.~Haddon, D.~Murphy, S.~Glarum, T.~Palstra, A.~Ramirez, A.~Karton, Nature 350 (1991) 600--601.

\bibitem{yonezawa2004}
S.~Yonezawa, Y.~Muraoka, Z.~Hiroi, Journal of the physical society of japan 73 (2004) 1655--1656.

\bibitem{hiroi2011}
Z.~Hiroi, J.~I. Yamaura, K.~Hattori, Journal of the physical society of japan 81 (2011) 011012.

\bibitem{hiroi2004}
Z.~Hiroi, S.~Yonezawa, Y.~Muraoka, Journal of the physical society of japan 73 (2004) 1651--1654.

\bibitem{saniz2004}
R.~Saniz, J.~E. Medvedeva, L.~H. Ye, T.~Shishidou, A.~J. Freeman, Physical review b 70 (2004) 100505.

\bibitem{hiroi2005}
Z.~Hiroi, S.~Yonezawa, J.~I. Yamaura, T.~Muramatsu, Y.~Muraoka, Journal of the physical society of japan 74 (2005) 1682--1685.

\bibitem{muramatsu2004}
T.~Muramatsu, S.~Yonezawa, Y.~Muraoka, Z.~Hiroi, Journal of the physical society of japan 73 (2004) 2912--2913.

\bibitem{yamaura2006}
J.~I. Yamaura, S.~Yonezawa, Y.~Muraoka, Z.~Hiroi, Journal of solid state chemistry 179 (2006) 336--340.

\bibitem{saniz2005}
R.~Saniz, A.~J. Freeman, Physical review b 72 (2005) 024522.

\bibitem{yamanaka2000}
S.~Yamanaka, E.~Enishi, H.~Fukuoka, M.~Yasukawa, Inorganic chemistry 39 (2000) 56--58.

\bibitem{bouquet2001}
F.~Bouquet, R.~Fisher, N.~Phillips, D.~Hinks, J.~Jorgensen, Physical review letters 87 (2001) 047001.

\bibitem{aydemir2010}
U.~Aydemir, C.~Candolfi, H.~Borrmann, M.~Baitinger, A.~Ormeci, W.~Carrillo-Cabrera, C.~Chubilleau, B.~Lenoir, A.~Dauscher, N.~Oeschler, et~al., Dalton transactions 39 (2010) 1078--1088.

\bibitem{heil2019}
C.~Heil, S.~D. Cataldo, G.~B. Bachelet, L.~Boeri, Physical review b 99 (2019) 220502.

\bibitem{drozdov2019}
A.~Drozdov, P.~Kong, V.~Minkov, S.~Besedin, M.~Kuzovnikov, S.~Mozaffari, L.~Balicas, F.~Balakirev, D.~Graf, V.~Prakapenka, et~al., Nature 569 (2019) 528--531.

\bibitem{somayazulu2019}
M.~Somayazulu, M.~Ahart, A.~K. Mishra, Z.~M. Geballe, M.~Baldini, Y.~Meng, V.~V. Struzhkin, R.~J. Hemley, Physical review letters 122 (2019) 027001.

\bibitem{hong2020}
F.~Hong, L.~Yang, P.~Shan, P.~Yang, Z.~Liu, J.~Sun, Y.~Yin, X.~Yu, J.~Cheng, Z.~Zhao, Chinese physics letters 37 (2020) 107401.

\bibitem{snider2021}
E.~Snider, N.~Dasenbrock-Gammon, R.~McBride, X.~Wang, N.~Meyers, K.~V. Lawler, E.~Zurek, A.~Salamat, R.~P. Dias, Physical review letters 126 (2021) 117003.

\bibitem{guo2013}
J.~Guo, J.~I. Yamaura, H.~Lei, S.~Matsuishi, Y.~Qi, H.~Hosono, Physical review b 88 (2013) 140507.

\bibitem{ishida2014}
S.~Ishida, Y.~Yanagi, K.~Oka, K.~Kataoka, H.~Fujihisa, H.~Kito, Y.~Yoshida, A.~Iyo, I.~Hase, Y.~Gotoh, et~al., Journal of the american chemical society 136 (2014) 5245--5248.

\bibitem{duong2014}
H.~D. Nguyen, Y.~Prots, W.~Schnelle, B.~Boehme, M.~Baitinger, S.~Paschen, Y.~Grin, Zeitschrift für anorganische und allgemeine chemie 640 (2014) 760--767.

\bibitem{Zhao}
Y.~Zhao, J.~Deng, A.~Bhattacharyya, D.~Adroja, P.~Biswas, L.~Gao, W.~Cao, C.~Li, C.~Pei, T.~Ying, et~al., Chinese physics letters 38 (2021) 127402.

\bibitem{momma2008vesta}
K.~Momma, F.~Izumi, Journal of applied crystallography 41 (2008) 653--658.

\bibitem{Lee1999}
S.~L. Lee, R.~Cywinski, S.~Kilcoyne, Muon Science: Muons in Physics, Chemistry and Materials, Vol.~51, CRC Press, 1999.

\bibitem{sonier2000}
J.~E. Sonier, J.~H. Brewer, R.~F. Kiefl, Reviews of modern physics 72 (2000) 769.

\bibitem{Pratt2000}
F.~Pratt, Physica b: Condensed matter 289 (2000) 710--714.

\bibitem{kohn1965}
W.~Kohn, L.~J. Sham, Physical review 140 (1965) A1133.

\bibitem{dal2014}
A.~D. Corso, Computational materials science 95 (2014) 337--350.

\bibitem{perdew1996}
J.~P. Perdew, K.~Burke, M.~Ernzerhof, Physical review letters 77 (1996) 3865.

\bibitem{marzari1999}
N.~Marzari, D.~Vanderbilt, A.~D. Vita, M.~C. Payne, Physical review letter 82 (1999) 3296.

\bibitem{das2021probing}
D.~Das, D.~Adroja, M.~Lees, R.~Taylor, Z.~Bishnoi, V.~Anand, A.~Bhattacharyya, Z.~Guguchia, C.~Baines, H.~Luetkens, et~al., Physical review b 103 (2021) 144516.

\bibitem{zhao2021superconductivity}
Y.~Zhao, J.~Deng, A.~Bhattacharyya, D.~Adroja, P.~Biswas, L.~Gao, W.~Cao, C.~Li, C.~Pei, T.~Ying, et~al., Chinese physics letters 38 (2021) 127402.

\bibitem{bhattacharyya2018}
A.~Bhattacharyya, D.~Adroja, M.~Smidman, V.~Anand, Science china physics, mechanics \& astronomy 61 (2018) 1--22.

\bibitem{adroja2017}
A.~Bhattacharyya, M.~R. Lees, K.~Panda, P.~P. Ferreira, T.~T. Dorini, E.~Gaudry, L.~T.~F. Eleno, V.~K. Anand, J.~Sannigrahi, P.~K. Biswas, et~al., Phys. rev. mater. 6(6) (2022) 064802.

\bibitem{bhattacharyya2019}
A.~Bhattacharyya, D.~Adroja, K.~Panda, S.~Saha, T.~Das, A.~Machado, O.~Cigarroa, T.~Grant, Z.~Fisk, A.~Hillier, et~al., Physical review letters 122 (2019) 147001.

\bibitem{adroja2021pairing}
D.~Adroja, A.~Bhattacharyya, Y.~Sato, M.~Lees, P.~Biswas, K.~Panda, V.~Anand, G.~B. Stenning, A.~Hillier, D.~Aoki, Physical review b 103 (2021) 104514.

\bibitem{svanidze2015non}
E.~Svanidze, L.~Liu, B.~Frandsen, B.~White, T.~Besara, T.~Goko, T.~Medina, T.~Munsie, G.~Luke, D.~Zheng, et~al., Physical review x 5 (2015) 011026.

\bibitem{singh2014detection}
R.~P. Singh, A.~D. Hillier, B.~Mazidian, J.~Quintanilla, J.~Annett, D.~M. Paul, G.~Balakrishnan, M.~Lees, Physical review letters 112 (2014) 107002.

\bibitem{biswas2012comparative}
P.~Biswas, A.~Hillier, M.~Lees, D.~M. Paul, Physical review b 85 (2012) 134505.

\bibitem{bhattacharyya2019investigation}
A.~Bhattacharyya, K.~Panda, D.~Adroja, N.~Kase, P.~Biswas, S.~Saha, T.~Das, M.~Lees, A.~Hillier, Journal of physics: Condensed matter 32 (2019) 085601.

\bibitem{prozorov2006}
R.~Prozorov, R.~W. Giannetta, Superconductor science and technology 19 (2006) R41.

\bibitem{annett1990}
J.~F. Annett, Advances in physics 39 (1990) 83--126.

\bibitem{BCS}
J.~Bardeen, L.~N. Cooper, J.~R. Schrieffer, Physical review 106 (1957) 162.

\bibitem{lin2003bcs}
J.~Y. Lin, P.~Ho, H.~Huang, P.~Lin, Y.~L. Zhang, R.~C. Yu, C.~Q. Jin, H.~Yang, Physical review b 67 (2003) 052501.

\bibitem{mu2007possible}
G.~Mu, Y.~Wang, L.~Shan, H.~H. Wen, Physical review b 76 (2007) 064527.

\bibitem{sun2004magnetic}
C.~Sun, J.~Y. Lin, S.~Mollah, P.~Ho, H.~Yang, F.~Hsu, Y.~Liao, M.~Wu, Physical review b 70 (2004) 054519.

\bibitem{amato1997}
A.~Amato, Reviews of modern physics 69 (1997) 1119.

\bibitem{chia2004}
E.~E. Chia, M.~Salamon, H.~Sugawara, H.~Sato, Physical review b 69 (2004) 180509.

\bibitem{mcmillan1968}
W.~McMillan, Physical review 167 (1968) 331.

\bibitem{ferreira2018}
P.~Ferreira, F.~Santos, A.~Machado, H.~Petrilli, L.~Eleno, Physical review b 98 (2018) 045126.

\bibitem{bhattacharyya2019_2}
A.~Bhattacharyya, D.~Adroja, P.~Biswas, Y.~Sato, M.~Lees, D.~Aoki, A.~Hillier, Journal of physics: Condensed matter 32 (2019) 065602.

\bibitem{Errea2015}
I.~Errea, M.~Calandra, C.~J. Pickard, J.~Nelson, R.~J. Needs, Y.~Li, H.~Liu, Y.~Zhang, Y.~Ma, F.~Mauri, High-pressure hydrogen sulfide from first principles: A strongly anharmonic phonon-mediated superconductor, Physical Review Letters 114~(15) (2015).

\bibitem{Talantsev2020}
E.~F. Talantsev, Advanced mcmillan’s equation and its application for the analysis of highly-compressed superconductors, Superconductor Science and Technology 33~(9) (2020) 094009.

\bibitem{guo2013superconductivity}
J.~Guo, J.~i~Yamaura, H.~Lei, S.~Matsuishi, Y.~Qi, H.~Hosono, Physical review b 88 (2013) 140507.

\bibitem{berk1966}
N.~Berk, J.~Schrieffer, Physical review letters 17 (1966) 433.

\bibitem{fay1980}
D.~Fay, J.~Appel, Physical review b 22 (1980) 3173.

\bibitem{moriya2003}
T.~Moriya, K.~Ueda, Reports on progress in physics 66 (2003) 1299.

\bibitem{floris2007}
A.~Floris, A.~Sanna, S.~Massidda, E.~Gross, Physical review b 75 (2007) 054508.

\bibitem{bersier2009}
C.~Bersier, A.~Floris, A.~Sanna, G.~Profeta, A.~Continenza, E.~Gross, S.~Massidda, Physical review b 79 (2009) 104503.

\bibitem{flores2015}
J.~A. Flores-Livas, A.~Sanna, Physical review b 91 (2015) 054508.

\bibitem{heil2017}
C.~Heil, S.~Ponc´e, H.~Lambert, M.~Schlipf, E.~R. Margine, F.~Giustino, Physical review letter 119 (2017) 087003.

\bibitem{bhattacharyya2020}
A.~Bhattacharyya, P.~Ferreira, F.~Santos, D.~Adroja, J.~Lord, L.~Correa, A.~Machado, A.~Manesco, L.~T. Eleno, Physical review research 2 (2020) 022001.

\bibitem{zhao2020}
Y.~Zhao, C.~Lian, S.~Zeng, Z.~Dai, S.~Meng, J.~Ni, Physical review b 101 (2020) 104507.

\bibitem{de2021}
L.~R.~D. Faria, P.~P. Ferreira, L.~E. Correa, L.~T. Eleno, M.~S. Torikachvili, A.~J. Machado, Superconductor science and technology 34 (2021) 065010.

\bibitem{correa2021}
L.~E. Correa, P.~P. Ferreira, L.~R. de~Faria, T.~T. Dorini, M.~S. da~Luz, Z.~Fisk, M.~S. Torikachvili, L.~T. Eleno, A.~J. Machado, Journal of alloys and compounds 907 (2022) 164477.

\bibitem{kawamura2017}
M.~Kawamura, R.~Akashi, S.~Tsuneyuki, Physical review b 95 (2017) 054506.

\bibitem{pascut2019}
G.~L. Pascut, M.~Widom, K.~Haule, K.~F. Quader, Physical review b 100 (2019) 125114.

\bibitem{panda2019}
K.~Panda, A.~Bhattacharyya, D.~T. Adroja, N.~Kase, P.~K. Biswas, S.~Saha, T.~Das, M.~R. Lees, A.~D. Hillier, Physical review b 99(17) (2019) 174513.

\bibitem{bhattacharyya2021}
A.~Bhattacharyya, P.~P. Ferreira, K.~Panda, S.~H. Masunaga, L.~R. de~Faria, L.~Corre, F.~Santos, D.~Adroja, K.~Yokoyama, T.~Dorini, et~al., Journal of physics: Condensed matter 34(3) (2021) 035602.

\end{thebibliography}
  
\end{document}